\documentclass[twocolumn,conference]{IEEEtran}
\usepackage[T1]{fontenc}
\usepackage{units}
\usepackage{amsmath}
\usepackage{amssymb}
\usepackage{graphicx}

\makeatletter
\usepackage[caption=false,font=footnotesize]{subfig}

\usepackage[keeplastbox]{flushend}

\IEEEoverridecommandlockouts

\makeatother

\begin{document}

\title{A Normalization Model for Analyzing Multi-Tier Millimeter Wave Cellular
Networks}

\author{\IEEEauthorblockN{Siqing Xiong\IEEEauthorrefmark{1}, Lijun Wang\IEEEauthorrefmark{2}\IEEEauthorrefmark{3},
Kyung Sup Kwak\IEEEauthorrefmark{4}, Zhiquan Bai\IEEEauthorrefmark{5},
Jiang Wang\IEEEauthorrefmark{6}\IEEEauthorrefmark{7}\IEEEauthorrefmark{8},
Qiang Li\IEEEauthorrefmark{1} and Tao Han\IEEEauthorrefmark{1}}\IEEEauthorblockA{\IEEEauthorrefmark{1}School of Electronic Information and Communications,
Huazhong University of Science and Technology, Wuhan, China}\IEEEauthorblockA{\IEEEauthorrefmark{2}School of Electronic Information, Wuhan University,
Wuhan, China}\IEEEauthorblockA{\IEEEauthorrefmark{3}Faculty of information science and technology,
Wenhua College, Wuhan, China}\IEEEauthorblockA{\IEEEauthorrefmark{4}Inha Hanlim Fellow Professor, Department of
Information and Communication, Inha university, Incheon, Korea}\IEEEauthorblockA{\IEEEauthorrefmark{5}School of Information Science and Engineering,
Shandong University, Jinan, China}\IEEEauthorblockA{\IEEEauthorrefmark{6}Shanghai Research Center for Wireless Communications,
Shanghai, China}\IEEEauthorblockA{\IEEEauthorrefmark{7}Shanghai Institute of Microsystem and Information
Technology, Chinese Academy of Sciences, Shanghai, China}\IEEEauthorblockA{\IEEEauthorrefmark{8}Key lab of wireless sensor network and communication,
Chinese Academy of Sciences, Shanghai, China}\IEEEauthorblockA{Email: \IEEEauthorrefmark{1}\{xiongsiqing, qli\_patrick, hantao\}@hust.edu.cn,
\IEEEauthorrefmark{2}\IEEEauthorrefmark{3}wanglijun@whu.edu.cn,
\IEEEauthorrefmark{4}kskwak@inha.ac.kr}\IEEEauthorblockA{\IEEEauthorrefmark{5}zqbai@sdu.edu.cn, \IEEEauthorrefmark{6}\IEEEauthorrefmark{7}\IEEEauthorrefmark{8}jiang.wang@wico.sh}\thanks{The corresponding author is Lijun Wang (e-mail: wanglijun@whu.edu.cn).}}
\maketitle
\begin{abstract}
Based on the distinguishing features of multi-tier millimeter wave
(mmWave) networks such as different transmit powers, different directivity
gains from directional beamforming alignment and path loss laws for
line-of-sight (LOS) and non-line-of-sight (NLOS) links, we introduce
a normalization model to simplify the analysis of multi-tier mmWave
cellular networks. The highlight of the model is that we convert a
multi-tier mmWave cellular network into a single-tier mmWave network,
where all the base stations (BSs) have the same normalized transmit
power 1 and the densities of BSs scaled by LOS or NLOS scaling factors
respectively follow piecewise constant function which has multiple
demarcation points. On this basis, expressions for computing the coverage
probability are obtained in general case with beamforming alignment
errors and the special case with perfect beamforming alignment in
the communication. According to corresponding numerical exploration,
we conclude that the normalization model for multi-tier mmWave cellular
networks fully meets requirements of network performance analysis,
and it is simpler and clearer than the untransformed model. Besides,
an unexpected but sensible finding is that there is an optimal beam
width that maximizes coverage probability in the case with beamforming
alignment errors.
\end{abstract}


\section{Introduction}

Given the dearth of spectrum in sub-3GHz bands, use of higher frequency
bands is indispensable to meet the projected data demands of 2020
\cite{Cisco2013Whitepaper}. Faced with this challenge, cellular systems
based on the millimeter wave (mmWave) bands has been attracted lots
of interest, between 30 and 300 GHz, where the available bandwidths
are much wider than today's cellular networks \cite{ZPi2011CM,Rangan2014JPROC}.
Recent field measurements also reveal the prospect of mmWave signals
for the access link between the user equipment (UE) and base station
(BS) in cellular systems \cite{Rappaport2013TAP}.

Evaluating the system performance of mmWave cellular networks is a
crucial task in order to understand the network behavior. Recently,
several studies analyze the coverage performance and capacity in mmWave
cellular networks using results from stochastic geometry \cite{Bai2014MCOM,Kulkarni2014GLOCOM}.
In \cite{Singh2014ACSSC}, a tractable model is proposed for user's
rate distribution in noise-limited mmWave cellular networks, and a
general framework has been proposed to evaluate coverage performance
of the mmWave networks in \cite{Bai2015TWC}. However, one must remember
that the mmWave cellular communication is easily affected by propagation
environmental factors such as atmospheric conditions and physical
obstacles, so the analysis of system-level performance evaluation
of mmWave cellular network is usually single-tier network. \cite{Ghosh2012MCOM}
shows that cellular networks are becoming less regular as a variety
of demand-based low power nodes are being deployed, and small cell
networks were studied in recently literature \cite{Ge2014Network,Ge2015IET,Ge2016WCom}.
Therefore, the networks could be regarded as the multi-tier cellular
networks instead of the simple single-tier network. Moreover, as one
of the candidate technologies in 5G, mmWave will be widely applied
to various BSs with different transmit powers, antenna gains, etc
\cite{Rangan2014JPROC}. Therefore, the emergence of multi-tier mmWave
cellular networks is inevitable. 

Recently a few researchers have presented some initial analysis of
multi-tier mmWave cellular networks with the aid of stochastic geometry
\cite{Renzo2015TWC}. However, the mathematical framework for multi-tier
mmWave cellular networks is not clear, and currently available mathematical
framework presented on \cite{Dhillon2012JSAC,Ge2015TCOM} for modeling
micro wave cellular networks is not directly applicable to mmWave
cellular networks. The main reasons are related to the need of incorporating
realistic path-loss, blockage models and highly directional antenna
gains. They are significantly different from micro wave communications.
The investigations \cite{Rangan2014JPROC,Singh2015JSAC} have demonstrated
large bandwidth mmWave networks tend to be noise-limited in urban
settings with blocking, in contrast to micro wave cellular networks,
which are interference-limited. Therefore, a tractable model for characterizing
the multi-tier mmWave cellular networks seems important to develop.

In this paper, we aim at proposing a normalization model which can
simplify analysis of multi-tier mmWave cellular networks. To the best
of our knowledge, the works converting multi-tier networks into single-tier
network in mmWave communication systems have not yet been analyzed
until now. Moreover, we derived the end-to-end signal-to-noise ratio
(SNR) in general case with beamforming alignment errors in the communication,
and discussed the SNR in perfect beamforming alignment case. The numerical
results proved that the normalization model is an effective model
for analysis of multi-tier mmWave cellular networks.

\section{System Model}

In this section, we introduce our system model for down-link multi-tier
mmWave cellular networks composed of $K$ independent network tiers
of BSs with different deployment densities, transmit powers, and antenna
gains. It is assumed that the BSs belonging to $k$-th tier are distributed
uniformly in $\mathbb{R}^{2}$ according to a bi-dimensional homogeneous
Poisson point process (PPP) $\Phi_{k}$ of density $\lambda_{k}$
, and have transmit power $P_{k}$ and the same beam width of the
main lobe $\omega\in\left(0,2\pi\right)$. Assuming that the multiple
cells of different tiers are distributed in the same plane, then,
the distribution of the BSs in multi-tier mmWave networks is defined
as $\Phi=\cup_{k=1}^{K}\Phi_{k}$ with density $\lambda=\sum_{k=1}^{K}\lambda_{k}$.
Without loss of generality, the typical UE is assumed to be located
at the origin $\left(0,0\right)$ and the distance between an arbitrary
BS and typical UE is $x$. 

\subsection{Directional Beamforming Model}

Antenna arrays are deployed at both BSs and UEs to perform directional
beamforming. For analytical tractability, the actual antenna patterns
are approximated by a sectored antenna model. The simple model captures
the interplay between the antenna gain and half-power beam width.
Let $G_{q}\left(\theta\right)$ be an ideal sector antenna with beam
width $\omega$, main beam gain $M_{q}$, and side lobe gain $m_{q}$
with $0\leq m_{q}<1<M_{q}$. In particular, the antenna gains of a
generic BSs and UEs are denoted by $G_{\mathrm{BS}}\left(\theta\right)$,
$G_{\mathrm{UE}}\left(\theta\right)$, respectively, and $\theta$
is the angle off the boresight direction. That is
\begin{equation}
G_{q}\left(\theta\right)=\begin{cases}
M_{q}=\frac{2\pi-\left(2\pi-\omega\right)\epsilon}{\omega}, & \mathrm{if}\begin{array}{cc}
 & \left|\theta\right|\leq\frac{\omega}{2}\end{array}\\
m_{q}=\epsilon, & \mathrm{Otherwise}
\end{cases},\label{eq:gain}
\end{equation}
where $q\in\left\{ \mathrm{BS,UE}\right\} $, $\epsilon\ll1$. Let
$a_{j}=G_{\mathrm{BS}}\left(\theta\right)G_{\mathrm{UE}}\left(\theta\right)$
be the total directivity gain which is from BSs to the typical UE.
Considering the general situation, the errors in channel estimation
are not neglected, so the UE and serving BS have four directivity
gains based on beamforming alignment case. That is

\begin{equation}
a_{j}=\begin{cases}
M_{\mathrm{BS}}M_{\mathrm{UE}}, & \mathrm{with}\begin{array}{c}
j=1\end{array}\\
M_{\mathrm{BS}}m_{\mathrm{UE}}, & \mathrm{with}\begin{array}{c}
j=2\end{array}\\
m_{\mathrm{BS}}M_{\mathrm{UE}}, & \mathrm{with}\begin{array}{c}
j=3\end{array}\\
m_{\mathrm{BS}}m_{\mathrm{UE}}, & \mathrm{with}\begin{array}{c}
j=4\end{array}
\end{cases},\label{eq:alignment state}
\end{equation}
where $j\in\left\{ 1,2,3,4\right\} $ is referred to the beamforming
alignment state of the UE and serving BS. For example, if the main
lobe of beam between the UE and serving BS is alignment, the directivity
gain for the desired signal link is expressed as $a_{1}=M_{\mathrm{BS}}M$$_{\mathrm{UE}}$.

\subsection{Blockage Model}

Considering the characteristic of the mmWave, a BS with mmWave can
be either the line-of-sight (LOS) BS or the non-line-of-sight (NLOS)
BS to the typical UE, which is determined by the LOS probability function
$p_{los}\left(x\right)$. We adopted the blockage model proposed in
\cite{Singh2015JSAC} as an approximation of the statistical blockage
model \cite{Renzo2015TWC}, since it is simple yet flexible enough
to capture blockage statistics, and describe the coverage and rate
trends in mmWave cellular networks. The probability that a link length
$x$ is LOS is

\begin{equation}
p_{los}\left(x\right)=\begin{cases}
C, & \mathrm{if}\begin{array}{cc}
x\leq d\end{array}\\
0, & \mathrm{Otherwise}
\end{cases},\label{eq:los probability}
\end{equation}
where $0\leq C\leq1$. The parameters $\left(C,d\right)$ are geography
and deployment dependent. $C$ should be regarded as the average fraction
of LOS area in the circle of radius $d$ around the typical UE. Also,
the NLOS probability of a link is $1-p_{los}\left(x\right)$. To simplify
the analysis, we regard the circle of radius $d$ as a LOS  circle.

\subsection{SNR Model}

Recent studies on mmWave networks \cite{Singh2014ACSSC,Singh2015JSAC}
reveal that mmWave networks in urban settings are more noise limited,
in contrast to micro wave cellular networks, which are strongly interference-limited.
This is due to blocking sensitivity, the signals received from other
non-serving BSs can be almost negligible. Moreover, because the SNR
provides a good enough approximation to signal to interference plus
noise ratio (SINR) for directional mmWave cellular networks, we adopt
it to help us in derivation. 

The received power in the down-link at the typical UE from the serving
BS at location $\mathbf{x}$ is given as $p_{k}h_{x}a_{j}L\left(\mathbf{x}\right)$.
Here, $p_{k}$ represents the transmit power of $k$-th tier BSs,
and we assume independent Rayleigh fading for each link, the random
variable $h$ follows an exponential distribution with mean $\nicefrac{1}{\mu}$,
which is denoted as $h_{x}\sim\exp\left(\mu\right)$. $L\left(\mathbf{x}\right)=\left\Vert \mathbf{x}\right\Vert ^{-\alpha}$
is the pathloss, and $\alpha$ is the pathloss exponent. If the link
is LOS, $\alpha$ equals to $\alpha_{L}$ and $\alpha_{N}$ otherwise,
$\alpha_{L}<\alpha_{N}$. Then, the SNR at the typical UE from its
associated BS can be expressed as

\begin{equation}
SNR=\frac{p_{k}h_{x}a_{j}L\left(\mathbf{x}\right)}{N},
\end{equation}
where $N$ is the noise power.

\section{Normalization Model of Multi-tier mmWave Networks}

Different from the single-tier mmWave cellular network where all BSs
have the same transmit power, beam width and the main lobe gain, in
multi-tiers mmWave cellular networks, the BSs of different tiers have
different parameters in power, beam width and follow different distributions
geographically. The complexity of the scenario leads to many difficulties
and enormous computing work in the performance analysis. In order
to make the analysis clearer, we propose a normalization model in
this paper, which converts a multi-tier mmWave cellular networks to
a virtual single-tier cellular network by the method of scaling. Unlike
our previous works \cite{Han2015}, which proposed a transmission
power normalization model for conventional multi-tier heterogeneous
cellular networks, our model in this paper takes LOS/NLOS paths and
directional beamforming into consideration, which are critical factors
to mmWave networks. As a result, all BSs have the same normalized
transmission power 1, and then, the virtual distance and link type
which is either LOS or NLOS become two important factors that affect
the UE's received power.

In a $K$-Tier mmWave networks, the BSs in tier $k$, $k\in\left\{ 1,2,\cdots,K\right\} $,
have transmit power $p_{k}$, beam width of the main lobe $\omega_{k}$
and follow homogeneous PPP $\Phi_{k}$ of density $\lambda_{k}$.
Due to the characteristic of mmWave communication link, we assume
each tier network is split into two parts. Let $\Phi_{k}^{L}$ be
the point process of $k$-th tier LOS BSs, and $\Phi_{k}^{N}=\Phi_{k}\setminus\Phi_{k}^{L}$
be the point process of $k$-th tier NLOS BSs. The distribution of
all BSs in whole networks can be described as $\Phi=\cup_{k=1}^{K}\Phi_{k}=\cup_{k=1}^{K}\left(\Phi_{k}^{L}+\Phi_{k}^{N}\right)$.
Then, we will discuss separately the normalization model in two parts
(scaled by LOS factors and scaled by NLOS factors).

The received signal power at the typical UE from the BS at $\mathbf{x}\in\Phi_{k}$
is given as
\begin{align}
p_{kj} & =p_{k}a_{j}L\left(\mathbf{x}\right)h_{x}=1\cdot\left(\left(p_{k}a_{j}\right)^{-\frac{1}{\alpha}}\left\Vert \mathbf{x}\right\Vert \right)^{-\alpha}h_{x}\nonumber \\
 & =1\cdot\left\Vert \left(p_{k}a_{j}\right)^{-\frac{1}{\alpha}}\cdot\mathbf{x}\right\Vert ^{-\alpha}h_{x}=1\cdot L\left(\left(p_{k}a_{j}\right)^{-\frac{1}{\alpha}}\cdot\mathbf{x}\right)h_{x},\label{eq:normalized transmit power}
\end{align}
where 1 is the normalized transmit power and $L\left(\mathbf{x}\right)$
is the path loss function, and $j\in\left\{ 1,2,3,4\right\} $ indicates
that the user is associated with LOS BS in different four cases. For
example, $p_{k1}$ is the received signal power of typical UE in the
case where the main lobe of UE and serving BS are aligned. 

From \eqref{eq:normalized transmit power}, it is observed that the
signal power received at the typical UE located at $\left(0,0\right)$
from the BS at $\mathbf{x}$ is equal to that received from the virtual
BS with transmit power 1 and located at $\mathbf{x}_{L}=\left(p_{k}a_{j}\right)^{-\frac{1}{\alpha_{L}}}\cdot\mathbf{x}$.
Based on this, the $k$-th tier homogeneous PPP $\Phi_{k}$ can be
respectively scaled to $\Phi_{kj}^{'}=\left(p_{k}a_{j}\right)^{-\frac{1}{\alpha_{L}}}\Phi_{k}$
of the scaled density $\lambda_{kj}^{'}=\left(\frac{1}{\left(p_{k}a_{j}\right)^{-\frac{1}{\alpha_{L}}}}\right)^{2}\lambda_{k}=\left(p_{k}a_{j}\right)^{\frac{2}{\alpha_{L}}}\lambda_{k}$.
Simultaneously, the radius $d$ of LOS circle will be scaled by different
factors in each tier network, the LOS probability function by LOS
scaling factors is given by
\begin{equation}
p_{los}\left(x_{L}\right)=\begin{cases}
C, & \mathrm{if}\begin{array}{cc}
 & x_{L}\leq d_{kj}\end{array}\\
0, & \mathrm{Otherwise}
\end{cases},
\end{equation}
where $d_{kj}=d\cdot\left(p_{k}a_{j}\right)^{-\frac{1}{\alpha_{L}}}$,
that is the scaled radius of LOS circle in $k$-th tier networks and
in four beamforming alignment cases. For every $j\in\left\{ 1,2,3,4\right\} $,
let $\mathcal{D}_{j}=\left\{ d_{kj},k\in\left\{ 1,\ldots,K\right\} \right\} $
denote the set of scaled radius of $k$-th tier networks LOS circle,
and suppose that the elements of $\mathcal{D}_{j}$ are indexed in
an increasing order, that is $d_{\upsilon\left(1\right)j}\leq d_{\upsilon\left(2\right)j}\leq\cdots\leq d_{\upsilon\left(K\right)j}$
, and define $\gamma_{ij}=d_{\upsilon\left(i\right)j}$ as the scaled
radius of LOS circle in the order list $\gamma_{j}=\left\{ \gamma_{1j},\ldots,\gamma_{Kj}\right\} $.
And the corresponding scaled density can be written as $\lambda_{\upsilon\left(i\right)j}^{'}=\left(p_{\upsilon\left(i\right)}a_{\upsilon\left(i\right)j}\right)^{\frac{2}{\alpha_{L}}}\lambda_{\upsilon\left(i\right)}$.
The densities of the $K$-tier networks scaled by the LOS factors
in LOS circle can be given as
\begin{align}
\lambda_{j}^{L} & =\underset{i=1}{\sum^{K}}\underset{l\in\mathcal{C^{C}}}{\sum}\mathnormal{\lambda_{\upsilon\left(l\right)j}^{'}\mathbb{I}\left(\gamma_{\left(i-1\right)j}<x_{L}\leq\gamma_{ij}\right)}\nonumber \\
 & \begin{array}{cc}
 & +0\cdot\mathnormal{\mathbb{I}\left(\gamma_{Kj}\leq x_{L}\right)}\end{array},\label{eq:normalization density}
\end{align}
where $\mathbb{I}\left(\cdot\right)$ is the indicator function, $\mathcal{C}=\left\{ 0,1,\ldots,i-1\right\} $
denotes the subset of the set $\mathrm{\mathcal{K}=}\left\{ 0,1,\ldots,K\right\} $,
its supplementary set is $\mathcal{C^{\mathcal{C}}}=\left\{ i,\ldots,K\right\} $,
and $\gamma_{0j}=0$. For ease of analysis, the second term of \eqref{eq:normalization density}
is the scaled density of LOS BS outside LOS circle. It is worth noting
that the densities in \eqref{eq:normalization density} are scaled
from the densities BSs in all tiers within the circle of radius $d$
including LOS BSs and NLOS BSs, and we can get the LOS BSs densities
in four beamforming alignment cases which are $C\cdot\lambda_{j}^{L}$
respectively. The normalization model scaled by LOS scaling factors
is equivalent to converting $K$-tier mmWave networks to the virtual
single-tier mmWave network, in which there are the same four beamforming
alignment cases and the scaled densities respectively follow piecewise
constant functions which have $K$ demarcation points in every case.
\begin{figure}[tbh]
\centering{}\includegraphics[width=9cm]{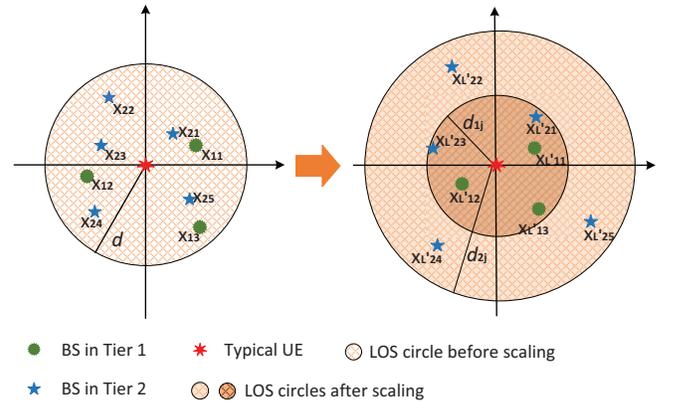}\caption{\label{fig:The-normalization-model}The normalization model of 2-tier
mmWave cellular networks in LOS case}
\end{figure}
 To explain this, consider a 2-tier mmWave network in LOS circle shown
in Fig. \ref{fig:The-normalization-model}. According to \eqref{eq:los probability},
the BSs located outside the circle of radius $d$ are NLOS BSs. In
Fig. \ref{fig:The-normalization-model}, BS $i$ is located at $\mathbf{x}_{ki}$
in $k$-th tier including BSs at $\mathbf{x}_{1i}$ in tier 1, and
BSs at $\mathbf{x}_{2i}$ in tier 2, with different transmit powers
and four directivity gains. The typical UE receives the desired signal
from the associated BS located at $\mathbf{x}_{ki}$. For ease of
analysis, we scale each tier by using different factors such that
virtual BSs at $\mathbf{x}_{L1i}^{'}$, $\mathbf{x}_{L2i}^{'}$ with
the same normalized power 1 are obtained, at the same time, the radius
$d$ of LOS circle is scaled to $d_{1j}$ and $d_{2j}$ by different
scaling factors so that the scaled density (BSs located in circle
of radius $d_{1j}$) is the sum of 2 tiers scaled density that is
$\lambda_{1j}^{'}+\lambda_{2j}^{'}$ as well as the scaled density
(BSs located at circular ring area between radius $d_{1j}$ and $d_{2j}$
) is $\lambda_{2j}^{'}$. Moreover, the probability of LOS BSs located
outside circle of radius $d_{2j}$ is 0, the scaled densities of BSs
in the LOS circle can be given as $\lambda_{j}^{L}=\left(\lambda_{1j}^{'}+\lambda_{2j}^{'}\right)\mathnormal{\mathbb{I}\left(0<x^{L}\leq d_{1j}\right)}$+$\lambda_{2j}^{'}\mathbb{I}\left(d_{ij}<x^{L}\leq d_{2j}\right)+0\cdot\left(d_{2j}<x^{L}\right)$.
It is worth emphasizing that the scaled densities respectively follow
piecewise constant functions which have two demarcation points in
four alignment cases and the probability of LOS BSs is $C$ in the
circle of radius $d_{2j}$, and the scaled densities of LOS BSs are
equal to $C\cdot\lambda_{j}^{L}$. 

It is known that the typical user does not directly communicate with
an NLOS BS, instead they communicate by radio wave's reflection and
scattering, etc, so that the directivity gains can be ignored. Assume
the typical user be associated with NLOS BS, according to \eqref{eq:normalized transmit power},
there is only a case $a_{j}=1$ for all $j$ in $\left\{ 1,2,3,4\right\} $.
And the signal power received at the typical UE from the BS at $\mathbf{x}$
is equal to that received from the virtual BS $\mathbf{x}_{N}=\left(p_{k}\right)^{-\frac{1}{\alpha_{N}}}\cdot\mathbf{x}$
with transmit power 1 and located at $\left(p_{k}\right)^{-\frac{1}{\alpha_{N}}}\cdot\mathbf{x}$.
Similarly, the $k$-th tier PPP $\Phi_{k}$ is scaled to $\Phi_{k}^{'}=\left(p_{k}\right)^{-\frac{1}{\alpha_{N}}}\Phi_{k}$
of the scaled density $\lambda_{k}^{'}=\left(p_{k}\right)^{\frac{2}{\alpha_{N}}}\lambda_{k}$.
The radius $d$ of LOS circle is scaled to $d_{k}^{'}=d\cdot p_{k}^{-\frac{1}{\alpha_{N}}}$,
and let $\mathcal{D}=\left\{ d_{k},k\in\left\{ 1,\ldots,K\right\} \right\} $
denote the set of scaled radius of $k$-th tier network LOS circle,
and suppose that the elements of $\mathcal{D}$ are indexed in an
increasing order, such that $d_{\upsilon\left(1\right)}\leq d_{\upsilon\left(2\right)}\leq\cdots\leq d_{\upsilon\left(K\right)}$
, and define $\gamma_{i}=d_{\upsilon\left(i\right)}$ as the scaled
radius of LOS circle in the order list $\gamma=\left\{ \gamma_{1},\ldots,\gamma_{K}\right\} $
, and the scaled density of BSs in $\upsilon\left(i\right)$-th tier
mmWave network is $\lambda_{\upsilon\left(i\right)}^{'}=\left(p_{\upsilon\left(i\right)}\right)^{\frac{2}{\alpha_{N}}}\lambda_{\upsilon\left(i\right)}$.
The density of the $K$-tier mmWave networks scaled by the NLOS scaling
factors in the LOS circle of radius $d$ with the NLOS probability
$\left(1-p_{los}\left(x\right)\right)$ can be given by

\begin{align}
\lambda_{1}^{N} & =\underset{i=1}{\sum^{K}}\underset{l\in\mathcal{C^{C}}}{\sum}\mathnormal{\lambda_{\upsilon\left(l\right)}^{'}\mathbb{I}\left(\gamma_{\left(i-1\right)}<x_{N}\leq\gamma_{i}\right)}\nonumber \\
 & \begin{array}{cc}
 & +0\cdot\mathnormal{\mathbb{I}\left(\gamma_{K}\leq x_{N}\right)},\end{array}\label{eq:nlos density}
\end{align}
where $\gamma_{0}=0$, and the scaled density of BSs outside the LOS
circle of radius $d$ with the NLOS probability 1 can be given by

\begin{align}
\lambda_{2}^{N} & =\underset{i=1}{\sum^{K}}\underset{l\in\mathcal{C}}{\sum}\mathnormal{\lambda_{\upsilon\left(l\right)}^{'}\mathbb{I}\left(\gamma_{\left(i-1\right)}<x_{N}\leq\gamma_{i}\right)}\nonumber \\
 & \begin{array}{cc}
 & +\underset{\mathit{{\scriptstyle l\in\mathrm{\mathcal{K}}}}}{\sum}\mathnormal{\lambda_{\upsilon\left(l\right)}^{'}\mathbb{I}\left(\gamma_{K}\leq x_{N}\right),}\end{array}\label{eq:nlos density two}
\end{align}
where $\lambda_{\upsilon\left(0\right)}^{'}=0$. From \eqref{eq:nlos density}
and \eqref{eq:nlos density two}, we can get $\lambda_{1}^{N}+\lambda_{2}^{N}=\underset{\mathit{{\scriptstyle l\in\mathcal{K}}}}{\sum}\mathnormal{\lambda_{\upsilon\left(l\right)}^{'}}$
that is scaled density of PPP $\Phi^{'}=\cup_{k=1}^{K}\Phi_{k}^{'}$
by the NLOS factors scaling. According to \eqref{eq:nlos density},
the scaled density of NLOS BSs in circle of radius $\gamma_{K}$ can
be expressed as $\left(1-C\right)\lambda_{1}^{N}$. Similarly, from
\eqref{eq:nlos density two}, the scaled density of NLOS BSs outside
the circle of radius $\gamma_{1}$ can be expressed as $1\cdot\lambda_{2}^{N}$.

\section{Coverage Analysis on the Normalization Model}

In Section IV, a general and tractable model for computing coverage
and rate of mmWave systems is provided, based on the assumptions of
$K$-tier mmWave cellular networks and a simple but flexible statistical
blockage model. With these assumptions, a virtual single-tier mmWave
network is presented with the new density of BSs. In this section,
assuming the typical UE is associated with the \textit{nearest} BS
in normalization model, so the typical UE can receive the maximum
signal power from that BS in $K$-tier mmWave cellular networks. And
then the expression of the coverage probability is provided for high-SNR
in general case (with beamforming alignment errors) and the special
case (with perfect beamforming alignment) based on the normalization
model. 

The coverage probability of $K$-tier mmWave cellular networks can
be given as 
\begin{equation}
\mathbf{\mathcal{P}_{cov}}\left(\mathrm{T,\omega}\right)=\mathbf{\mathbb{P}_{los}}+\mathbf{\mathbb{P}_{nlos}},\label{eq:sum}
\end{equation}
where $\mathbf{\mathbb{P}_{los}}$ is the probability when the typical
UE can be served by the \textit{nearest} BS with LOS path, $\mathbf{\mathbb{P}_{nlos}}$
is the probability when the typical UE can be served by the \textit{nearest}
BS with NLOS path. Then, we will respectively discuss the coverage
probabilities in two cases.

\subsection{The General Case}

Here, we will discuss the general case with beamforming alignment
errors, which is closer to the practical situation. Therefore, we
investigate the effect of beamforming alignment errors on coverage
probability. We employ an error model similar to that in \cite{Wildman2014TWC}.
Let $\left|\varepsilon_{q}\right|$ be the random additive beam-steering
errors, $q\in\left\{ \mathrm{BS,UE}\right\} $, $\varepsilon_{\mathrm{BS}}$
and $\varepsilon_{\mathrm{UE}}$ are independent of each other and
have a symmetrical distribution around $\omega_{q}$. In this paper,
we assume the beamwidth of main lobe $\omega_{\mathrm{BS}}=\omega_{\mathrm{UE}}=\omega$.
The probability density function (PDF) of the effective directivity
gain $a_{j}$ with beamforming alignment errors can be explicitly
written as \cite{Renzo2015TWC}
\begin{align}
 & f_{G}\left(a_{j}\right)=\nonumber \\
 & F_{\left|\varepsilon_{\mathrm{BS}}\right|}\left(\frac{\omega}{2}\right)F_{\left|\varepsilon_{\mathrm{UE}}\right|}\left(\frac{\omega}{2}\right)\delta\left(a_{j}-M_{\mathrm{BS}}M_{\mathrm{UE}}\right)\nonumber \\
 & +F_{\left|\varepsilon_{\mathrm{BS}}\right|}\left(\frac{\omega}{2}\right)\left(1-F_{\left|\varepsilon_{\mathrm{UE}}\right|}\left(\frac{\omega}{2}\right)\right)\delta\left(a_{j}-M_{\mathrm{BS}}m_{\mathrm{UE}}\right)\nonumber \\
 & +\left(1-F_{\left|\varepsilon_{\mathrm{BS}}\right|}\left(\frac{\omega}{2}\right)\right)F_{\left|\varepsilon_{\mathrm{UE}}\right|}\left(\frac{\omega}{2}\right)\delta\left(a_{j}-m_{\mathrm{BS}}M_{\mathrm{UE}}\right)\nonumber \\
 & +\left(1-F_{\left|\varepsilon_{\mathrm{BS}}\right|}\left(\frac{\omega}{2}\right)\right)\left(1-F_{\left|\varepsilon_{\mathrm{UE}}\right|}\left(\frac{\omega}{2}\right)\right)\delta\left(a_{j}-m_{\mathrm{BS}}m_{\mathrm{UE}}\right),\label{eq:alignment error}
\end{align}
where $\delta\left(\cdot\right)$ is the Kronecker's delta function,
$F_{\left|\epsilon_{q}\right|}\left(x\right)=\mathbf{\mathbb{P}}\left\{ \left|\varepsilon_{q}\right|\leq x\right\} $
is the cumulative distribution function of misalignment error. Assume
the beam-steering errors follow a Gaussian distribution with mean
equal to zero and variance equal to $\sigma_{\mathrm{BE}}^{2}$, so
absolute error $\left|\varepsilon\right|$ follows a half normal distribution
and $F_{\left|\varepsilon\right|}\left(x\right)=\mathrm{erf}\left(x/\left(\sqrt{2}\sigma_{\mathrm{BE}}\right)\right)$,
where $\mathrm{erf}\left(\cdot\right)$ denotes the error function.
From \eqref{eq:alignment error}, the probability that the typical
UE can be served by the \textit{nearest} BS with LOS path can be calculated
as 
\begin{align}
\mathbf{\mathbb{P}_{los}} & =\Pr\left(SNR_{\mathbf{los}}>\mathrm{T}\right)\nonumber \\
 & =\Pr\left(\frac{1\cdot h_{x}x_{L}^{-\alpha_{L}}}{N}>\mathrm{T}\right)\nonumber \\
 & =\int_{0}^{\infty}f_{G}\left(a_{j}\right)\mathrm{exp}\left(-\mathrm{T}Nx_{L}^{\alpha_{L}}\right)f_{j}^{L}\left(x_{L}\right)dx_{L}\nonumber \\
 & =\sum_{j=1}^{4}f_{G}\left(a_{j}\right)\int_{0}^{\infty}\mathrm{exp}\left(-\mathrm{T}Nx_{L}^{\alpha_{L}}\right)f_{j}^{L}\left(x_{L}\right)dx_{L},\label{eq:los cov pro in error}
\end{align}
where $f_{j}^{L}\left(x\right)=C\cdot2\pi x\lambda_{j}^{L}\mathrm{exp}\left(-\pi x^{2}\lambda_{j}^{L}\right)$.
And the probability that the typical UE can be served by the \textit{nearest}
BS with NLOS path can be given as

\begin{align}
\mathbf{\mathbb{P}_{nlos}} & =\Pr\left(SNR_{\mathbf{nlos}}>\mathrm{T}\right)\nonumber \\
 & =\Pr\left(\frac{1\cdot h_{x}x_{N}^{-\alpha_{N}}}{N}>\mathrm{T}\right)\nonumber \\
 & =\int_{0}^{\infty}\mathrm{exp}\left(-\mathrm{T}Nx_{N}^{\alpha_{N}}\right)f_{1}^{N}\left(x_{N}\right)dx_{N}\nonumber \\
 & \begin{array}{cc}
 & +\int_{0}^{\infty}\mathrm{exp}\left(-\mathrm{T}Nx_{N}^{\alpha_{N}}\right)f_{2}^{N}\left(x_{N}\right)dx_{N}\end{array},\label{eq:nlos coverage probability}
\end{align}
where $f_{1}^{N}\left(x\right)=\left(1-C\right)2\pi x\lambda_{1}^{N}\mathrm{exp}\left(-\pi x^{2}\lambda_{1}^{N}\right)$
and $f_{2}^{N}\left(x\right)=2\pi x\lambda_{2}^{N}\mathrm{exp}\left(-\pi x^{2}\lambda_{2}^{N}\right)$.
According to \eqref{eq:sum}, \eqref{eq:los cov pro in error} and
\eqref{eq:nlos coverage probability}, the coverage probability with
beamforming alignment errors between typical UE and serving BS can
be calculated.

\subsection{The Case with Perfect Beamforming Alignment }

In this part, we assume perfect beamforming alignment case, and obtain
the upper limit expression of coverage probability. Without beamforming
alignment errors, there is only a case where the maximum directivity
gain can be exploited on the intended link. According to \eqref{eq:alignment error},
we can get the PDF of the effective directivity gain $a_{j}$ in a
special case, that is $f_{G}\left(a_{j}\right)=1\cdot\delta\left(a_{j}-M_{\mathrm{BS}}M_{\mathrm{UE}}\right)+0\cdot\delta\left(a_{j}-M_{\mathrm{BS}}m_{\mathrm{UE}}\right)+0\cdot\delta\left(a_{j}-m_{\mathrm{BS}}M_{\mathrm{UE}}\right)+0\cdot\delta\left(a_{j}-m_{\mathrm{BS}}m_{\mathrm{UE}}\right)$.
Similarly, the probability that the typical UE is served by the \textit{nearest}
BS with LOS path can be presented as
\begin{align}
\mathbf{\mathbb{P}_{los}} & =\Pr\left(SNR_{\mathbf{los}}>\mathrm{T}\right)\nonumber \\
 & =\int_{0}^{\infty}f_{G}\left(a_{j}\right)\mathrm{exp}\left(-\mathrm{T}Nx_{L}^{\alpha_{L}}\right)f_{1}^{L}\left(x_{L}\right)dx_{L}\nonumber \\
 & =\int_{0}^{\infty}\mathrm{exp}\left(-\mathrm{T}Nx_{L}^{\alpha_{L}}\right)f_{1}^{L}\left(x_{L}\right)dx_{L},\label{eq:prefect alignment}
\end{align}
where $f_{1}^{L}\left(x\right)=C\cdot2\pi x\lambda_{1}^{L}\mathrm{exp}\left(-\pi x^{2}\lambda_{1}^{L}\right)$. 

According to \eqref{eq:nlos coverage probability} and \eqref{eq:prefect alignment},
the coverage probability can be expressed as

\begin{align}
\mathbf{\mathcal{P}_{cov}\left(\mathrm{T,\omega}\right)=} & \int_{0}^{\infty}\mathrm{exp}\left(-\mathrm{T}Nx_{L}^{\alpha_{L}}\right)f_{1}^{L}\left(x_{L}\right)dx_{L}\nonumber \\
 & +\int_{0}^{\infty}\mathrm{exp}\left(-\mathrm{T}Nx_{N}^{\alpha_{N}}\right)f_{1}^{N}\left(x_{N}\right)dx_{N}\nonumber \\
 & +\int_{0}^{\infty}\mathrm{exp}\left(-\mathrm{T}Nx_{N}^{\alpha_{N}}\right)f_{2}^{N}\left(x_{N}\right)dx_{N}.
\end{align}

\section{Numerical Results}

In this section, we explore the relationship between beam width and
maximum coverage probability with beamforming alignment errors. And
without loss of generality, we present some simulation results for
illustrating the normalization model and characterizing the coverage
performance of the 2-tier mmWave networks as well as the effect of
different network parameters. In all figures, LOS and NLOS path loss
exponents are $\alpha_{L}=2$ and $\alpha_{N}=4$, respectively.

\begin{figure}[tbh]
\begin{centering}
\includegraphics[width=9cm]{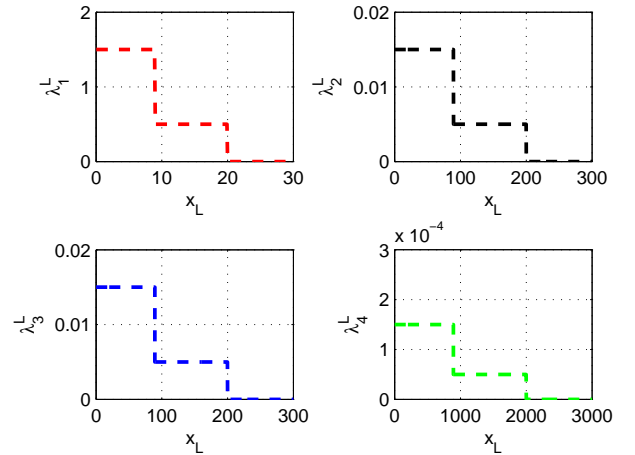}\caption{\label{fig:Normalization-densities(LOS)-of}The densities of 2-tier
mmWave networks scaled by the LOS scaling factors in four beamforming
alignment cases ($\alpha_{L}=2$, $M=10\mathrm{dB}$, $m=-10\mathrm{dB}$,
$p_{1}=1\mathrm{w}$, $p_{2}=5\mathrm{w}$, $\lambda_{1}=\frac{1}{200}$,
$\lambda_{2}=\frac{1}{500}$, $d=200\mathrm{m}$)}
\par\end{centering}
\end{figure}

In order to characterize the model clearer, the scaled densities which
are converted from 2-tier mmWave cellular networks to single-tier
network are shown in Fig. \ref{fig:Normalization-densities(LOS)-of}.
There are four scaled densities in different beamforming alignment
state of \ref{eq:alignment state}. In Fig. \ref{fig:Normalization-densities(LOS)-of},
the 2nd graph and the 3rd graph is the same, because we assume the
same beam width of the main lobe between user and BSs, the directivity
gains in the cases $j=2$ and case $j=3$ are the same. The scaled
densities follow piecewise constant functions. The piecewise points
are affected by directivity gains and transmit powers, respectively,
which provide convenience for performance analysis in multi-tier mmWave
cellular networks. And the densities scaled by NLOS scaling factors
are similar to that.

\begin{figure}[tbh]
\centering{}\includegraphics[width=8.5cm]{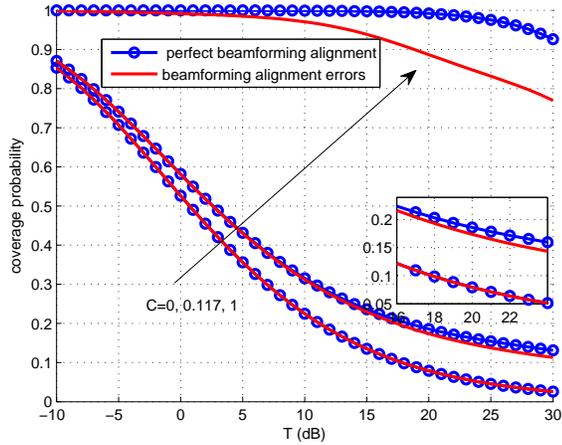}\caption{\label{fig:The-comparison-of}The comparison of coverage probability
between perfect beamforming alignment and beamforming alignment errors
cases ($\alpha_{L}=2$, $\alpha_{N}=4$, $p_{1}=1\mathrm{w}$, $p_{2}=5\mathrm{w}$,
$\lambda_{1}=\frac{1}{200}$, $\lambda_{2}=\frac{1}{500}$, $\omega_{q}=20^{\circ}$,
$\sigma_{BE}=4^{\circ}$,$d=200\mathrm{m}$)}
\end{figure}

In Fig. \ref{fig:The-comparison-of}, we compare coverage probability
based on different $\left(C,d\right)$ pairs in perfect beamforming
alignment and beamforming alignment errors cases. The empirical $\left(C,d\right)$
pair for Manhattan is $\left(0.117,200\right)$ \cite{Singh2015JSAC}.
In addition, two special cases with LOS $\left(C=1\right)$ and NLOS
$\left(C=0\right)$ in the inner circle of radius $d=200$ are considered,
and it can be interpreted to the upper limit and lower limit of coverage
probability in beamforming alignment errors cases. From the Fig. \ref{fig:The-comparison-of},
the coverage probability under the condition of perfect beamforming
alignment is higher than that in beamforming alignment errors case
at the same threshold. However, as the LOS probability $C$ decreases,
the difference becomes smaller until it tends to zero.

\begin{figure}[tbh]
\centering{}\includegraphics[width=9cm]{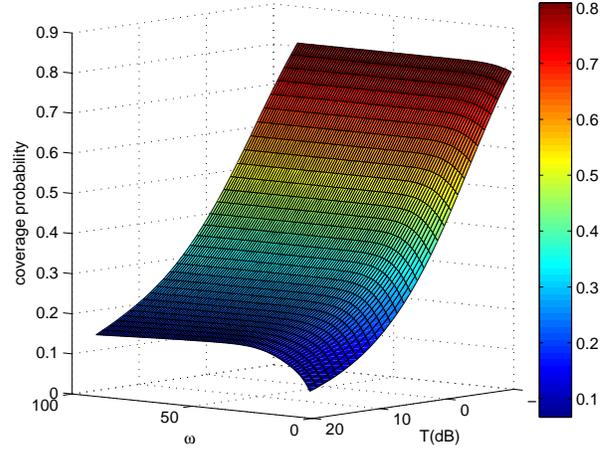}\caption{\label{fig:the-coverage-probability}The coverage probability with
beamwidth changing in different threshold ($\alpha_{L}=2$,$\alpha_{N}=4$,$p_{1}=1\mathrm{w}$,$p_{2}=5\mathrm{w}$,$\lambda_{1}=\frac{1}{200}$,
$\lambda_{2}=\frac{1}{500}$, $\sigma_{BE}=4^{\circ}$,$\left(C,d\right)=\left(0.117,200\right)$)}
\end{figure}

In Fig. \ref{fig:the-coverage-probability}, the effect of beam widths
on the coverage performance is analyzed in different SNR threshold.
It is considered that the UE's beam width is equal to the BS's beam
width and the side lobe strength $m_{q}=\epsilon$ is fixed in simulation.
From Fig. \ref{fig:the-coverage-probability}, we can see that there
exists a beam width to meet the highest coverage probability at the
same threshold. And it can be interpreted that the smaller the beam
width is, the greater the main lobe gain and the beamforming alignment
errors are, so that there will be the beam width which makes the coverage
probability maximum at the same threshold.

\section{Conclusions}

In this paper, a normalization model for simplifying analysis and
computation of multi-tier mmWave cellular networks was introduced.
Its novelty lies in converting the multi-tier mmWave cellular networks
into a virtual single-tier mmWave network, where all BSs have the
same normalized transmit power 1 and the scaled densities respectively
follow corresponding piecewise constant functions. We have adopted
the proposed approach to analyzing some coverage performance and the
effect of beamforming alignment errors based on the noise-limited
mmWave cellular systems in this paper. Numerical simulations have
confirmed that the results met the analysis requirements. In future
work, it would be interesting to analyze more system performances
under the normalization model. 

\section*{Acknowledgements}

The authors would like to acknowledge the support of the International
Science and Technology Cooperation Program of China (Grant Nos. 2015DFG12580
and 2014DFA11640), the National Natural Science Foundation of China
(Grant Nos. 61471180, 61461136004, 61301147, 61471347, and 61210002),
the Hubei Provincial Department of Education Scientific Research Project
(Grant No. B2015188), the Shanghai Natural Science Foundation (Grant
No. 16ZR1435100), the Fundamental Research Funds of Shandong University
(Grant No. 2016JC010), a grant from Wenhua College (Grant No. 2013Y08),
the National Research Foundation of Korea-Grant funded by the Korean
Government (Ministry of Science, ICT and Future Planning)-NRF-2014K1A3A1A20034987),
the EU FP7-PEOPLE-IRSES (Grant No. 610524), and the EU H2020 project
(Grant No. 723227). This research is supported by the China International
Joint Research Center of Green Communications and Networking (No.
2015B01008).

\appendices{}

\bibliographystyle{IEEEtran}
\bibliography{normalization}

\end{document}